\documentclass[prd,a4paper]{revtex4}
\usepackage{epsfig}

\newcommand{\be}{\begin{equation}}
\newcommand{\ee}{\end{equation}}
\newcommand{\bea}{\begin{eqnarray}}
\newcommand{\eea}{\end{eqnarray}}
\newcommand{\bean}{\begin{eqnarray*}}
\newcommand{\eean}{\end{eqnarray*}}

\newcommand{\gapproxeq}{\lower
.7ex\hbox{$\;\stackrel{\textstyle >}{\sim}\;$}}
\newcommand{\lapproxeq}{\lower
.7ex\hbox{$\;\stackrel{\textstyle <}{\sim}\;$}}

\begin{document}

\bibliographystyle{unsrt}

\title{\bf Glueball-$Q\bar{Q}$ mixing and Okuba-Zweig-Iizuka rule violation 
in the hadronic decays of heavy quarkonia }

\author{Qiang Zhao$^{1,2}$\footnote{e-mail: qiang.zhao@surrey.ac.uk}, 
Bing-Song Zou$^1$\footnote{e-mail: zoubs@ihep.ac.cn}, 
and Zhong-Biao Ma$^3$
}

\affiliation{1) Institute of High Energy Physics,
Chinese Academy of Sciences, Beijing, 100049, P.R. China}

\affiliation{2)Department of Physics,
University of Surrey, Guildford, GU2 7XH, United Kingdom}

\affiliation{3) CCAST, Beijing 100080, P.R. China}

\date{\today}

\begin{abstract}

We investigate the correlations between the scalar meson configurations 
and Okuba-Zweig-Iizuka (OZI) rule violations 
in the hadronic decays of heavy quarkonia,  
e.g. $J/\psi$ and $\Upsilon$, into isoscalar vector meson ($\phi$ and $\omega$) 
and scalar mesons ($f_0(1710)$, $f_0(1500)$, and $f_0(1370)$). 
It shows that the dramatic change of the values of 
the branching ratio fraction of $\phi f_0/\omega f_0$ 
from low (e.g. in $J/\psi$ decays) 
to high energies (e.g. in $\Upsilon$ decays) 
will not only test the glueball-$Q\bar{Q}$ mixings, 
but also provide important information 
about the mysterious OZI-rule violations within the scalars.

\end{abstract}

\maketitle


The recent experimental data from BES~\cite{bes-plb,bes-phi,bes-f0(1500),jin} 
seem to have enhanced the possible 
existence of significant glueball contents within 
the low-lying scalars~\cite{scalar-meson-1,close-amsler,close-ichep04,mp,ukqcd,chanowitz,narison,cfl,bugg}.
In particular, large glueball components 
are preferred in the $f_0(1500)$, while large $|s\bar{s}\rangle$ 
and $|n\bar{n}\rangle\equiv |u\bar{u}+d\bar{d}\rangle/\sqrt{2}$ 
are dominant in the $f_0(1710)$ and $f_0(1370)$, respectively. 
Although the glueball components are relatively small in these two 
states, their phases relative to the dominant $Q\bar{Q}$ components 
turn out to play crucial roles 
in the understanding of the scalar 
spectroscopy~\cite{scalar-meson-1,close-amsler,close-kirk,close-zhao-f0}.  
Implications arising from the presence of the glueball components 
can then be detected in their production processes. 
Mechanisms such as $J/\psi$ hadronic and radiative decays are ideal 
for probing information about the glueball-$Q\bar{Q}$ 
mixings. 
 
\begin{figure}
\begin{center}
\epsfig{file=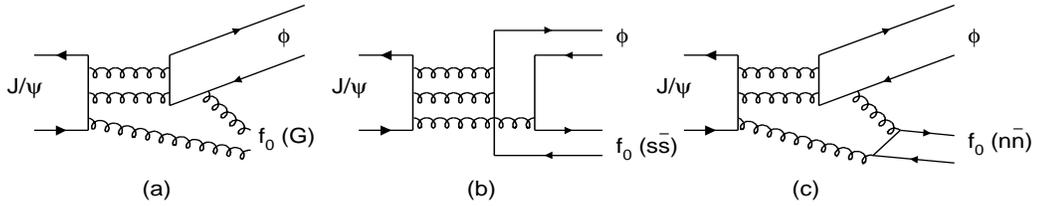, width=16cm,height=6.cm}
\caption{Schematic processes for the production 
of $f_0^i$ in $J/\psi\to\phi f_0^i$. 
}
\protect\label{fig-1}
\end{center}
\end{figure}

In Ref.~\cite{close-zhao-f0}, we reported a coherent study of the production 
of $f_0^i$ ($i=1$, 2, 3 corresponding to 
$f_0(1710)$, $f_0(1500)$, and $f_0(1370)$)  
in $J/\psi\to V f_0 \to V PP$, where $V=\phi, \ \omega$. 
There are  advantages of studying isoscalar vector meson $\phi$ and $\omega$ 
in association with the $f_0$ states: 
i) Since $\omega$ and $\phi$ are approximately ideally mixed 
in the flavor space, the process $J/\psi\to \phi f_0(s\bar{s})$ and 
$J/\psi\to\omega f_0(n\bar{n})$ may probe the $s\bar{s}$ and $n\bar{n}$ 
components separately to leading order via the singly OZI~\cite{ozi} disconnected 
diagrams (Fig.~\ref{fig-1}(b)); 
ii) If there exist glueball components in the $f_0$ states, 
those two processes also probe the glueball components 
at the same order via $J/\psi\to V G$ (Fig.~\ref{fig-1}(a)); 
iii) As discussed in Ref.~\cite{close-zhao-f0}, the ratio of these two decay 
channels also highlights the importance of the doubly OZI disconnected 
diagram in the interpretation of the BES data for the production of
$f_0(1370)$ and $f_0(1500)$ in the $J/\psi$ decays (see Fig.~\ref{fig-1}(c)), 
which is contrary to expectations based on naive OZI rule~\cite{seiden}.
Therefore, the importance of its violations can also be 
reflected by the failure of the prescriptions, which only include the 
leading contributions from the singly 
disconnected processes and the direct glueball production. 

It was pointed out by Lipkin before that the OZI-rule violations 
can always proceed by a two-step process involving 
intermediate virtual mesons~\cite{lipkin}. 
For example, as studied by Geiger and Isgur~\cite{isgur-geiger}, 
systematic cancellations among the hadronic loops occurred 
for the $u\bar{u}\leftrightarrow s\bar{s}$ mixing in all 
the low-lying nonets except $0^{++}$. A general argument 
was given by Lipkin and Zou~\cite{lipkin-zou}. 
Similar to the large rates of 
$p\bar{p}\to K^*\bar{K}+c.c.\to \phi\pi$~\cite{locher-lu-zou}, 
the doubly OZI disconnected processes in the $J/\psi$ decays, can occur via 
$J/\psi\to K^*\bar{K}+c.c.\to V f_0^i$, or 
$J/\psi\to \rho\pi+c.c.\to V f_0^i$. 
As we know that $J/\psi\to K^*\bar{K}+c.c.$ and $\rho\pi +c.c.$
are two of the largest decay channels of $J/\psi$~\cite{pdg2004}, 
the virtual meson mediated transitions as the dominant contributions 
to the doubly OZI disconnected processes may not be small. 
Such an approach was applied to the investigation of 
the ``$\rho\pi$ puzzle" in $J/\psi$ and $\psi^\prime$ 
by Li, Bugg and Zou~\cite{li-bugg-zou}. 
In comparison with the quark-gluon interaction picture 
the intermediate meson exchange model could be useful for us 
to study the correlation between the strong glueball-$Q\bar{Q}$ mixings and 
presence of possible OZI violations. In particular, 
the energy evolution of the doubly disconnected processes 
can be investigated and may shed light on some of those 
recently raised puzzles in the scalar meson productions~\cite{jin}.

As follows, we first summarize 
the glueball-$Q\bar{Q}$ mixing scheme and the factorizations for the 
$f_0^i$ productions in the quarkonium decays. 
By studying the exclusive reactions, 
$J/\psi\to V f_0^i\to VPP$, and $\Upsilon\to V f_0^i\to VPP$, where the 
``close-to-ideally-mixed" $\omega$ and $\phi$ play a role as an OZI filter, 
the different energy-dependence between the singly and doubly disconnected 
processes can be highlighted. We then study 
the energy evolution of the doubly OZI disconnected processes 
in an intermediate meson exchange model. The conclusion will be drawn 
in the end.

Applying the glueball-$Q\bar{Q}$ mixing scheme proposed by Amsler and 
Close~\cite{close-amsler}, and developed later 
by Close {\it et al}~\cite{close-kirk,close-zhao-f0}, the three $f_0^i$ states, 
which are the eigenstates of the glueball-$Q\bar{Q}$ mixing potential, 
can be described as
\be
\left(
\begin{array}{c}
|f_0(1710)\rangle \\
|f_0(1500)\rangle \\
|f_0(1370)\rangle 
\end{array}
\right) = 
\left(
\begin{array}{ccc}
x_1 & y_1 & z_1 \\
x_2 & y_2 & z_2 \\
x_3 & y_3 & z_3 
\end{array}
\right)
\left(
\begin{array}{c}
|G\rangle \\
|{s\bar{s}}\rangle \\
|{n\bar{n}}\rangle 
\end{array}
\right) \ ,
\ee
where $|G\rangle$, $|s\bar{s}\rangle$, and $|n\bar{n}\rangle$ are
the pure glueball and $Q\bar{Q}$ nonet, respectively, and 
$x_i$, $y_i$ and $z_i$ are the mixing matrix elements determined 
by the flavor-independent glueball-$Q\bar{Q}$ 
transitions~\cite{close-amsler,close-kirk}. 
As studied in Ref.~\cite{close-kirk,close-zhao-f0}, the mixing matrix 
determined by the data for 
$f_0^i\to PP$ possesses stable features and can be tested 
in the $f_0^i$ production processes. In particular, 
it was shown in Ref.~\cite{close-zhao-f0} that the description 
of $J/\psi\to Vf_0^i$ needs essentially a strong glueball 
component in the $f_0(1500)$. 
Furthermore, the preference of the presence of strong glueball 
component in the $f_0^i$ also implies the strong OZI violations 
could become significant at the low energy region.

In Ref.~\cite{close-zhao-f0}, an OZI-selecting factorization scheme
is proposed for $J/\psi\to V f_0^i$, and can be generalized to 
other vector quarkonium decays, e.g. $\Upsilon\to V f_0^i$. 
The transition occurs via  
\bea
M_{\phi G} &\equiv &\langle G | V_\phi |Q_V\rangle \nonumber\\
M_{\phi(s\bar{s})} &\equiv &\langle s\bar{s} |V_\phi|Q_V\rangle \nonumber\\
M_{\phi(n\bar{n})} &\equiv &\langle n\bar{n} |V_\phi|Q_V\rangle \ , 
\eea
where $Q_V$ denotes the quarkonium wavefunction, and 
$M_{\phi G}$, $M_{\phi (s\bar{s})}$, and $M_{\phi(n\bar{n})}$ 
are transition amplitudes for the production of the pure glueball and 
flavor nonet through the vector meson potential. Since the $\phi$ 
is almost $s\bar{s}$, the potential filters out the higher order 
OZI process $M_{\phi(n\bar{n})}$, which is doubly OZI disconnected 
at leading order.

Due to lack of information about the potentials, 
we assume, based on the order of $\alpha_s$, that the singly 
OZI disconnected processes have the same strengths. As an example 
for $J/\psi\to \phi f_0^i$, we assume 
\be 
\langle G|V_\phi |J/\psi\rangle
\simeq \langle s\bar{s}|V_\phi|J/\psi\rangle \ .
\ee
For the doubly disconnected transition 
$M_{\phi(n\bar{n})}$, it is natural to introduce $r$ for its 
relative strength to the singly disconnected ones:
\be
\label{para-r} 
\langle n\bar{n}|V_\phi|J/\psi\rangle 
\simeq \sqrt{2} r \langle s\bar{s}|V_\phi|J/\psi\rangle ,
\ee
where the factor $\sqrt{2}$ is from the normalization of 
$n\bar{n}$. Naively, $r<<1$ 
would imply that the doubly disconnected process has a perturbative 
feature, while $r \simeq 1$ reflects the strong OZI-rule violation. 
The same factorization can also be applied to 
$J/\psi\to \omega f_0^i$~\cite{close-zhao-f0}.

Following the assumption made for those three independent transitions,
and considering the partial decay widths for 
$J/\psi\to \phi f_0^i\to \phi PP$ and 
$J/\psi\to \omega f_0^i\to \omega PP$, 
we have a set of equations in terms of only one independent transition 
amplitude of the pure glueball production: 
\bea
\Gamma_{J/\psi\to\phi f_0^i\to\phi PP} & = & 
\frac{|{\bf p}_{\phi i}|}{|{\bf p}_{\phi G}|}
br_{f_0^i\to PP} [  x_i + y_i  + \sqrt{2}r z_i ]^2
\Gamma_{J/\psi\to \phi G} \ , \\
\Gamma_{J/\psi\to\omega f_0^i\to\omega PP} & = & 
\frac{|{\bf p}_{\omega i}|}{|{\bf p}_{\omega G}|}
br_{f_0^i\to PP} [ x_i + r y_i + \sqrt{2} z_i ]^2
\Gamma_{J/\psi\to \omega G} \ ,
\eea
where $2\Gamma_{J/\psi\to \phi G}/|{\bf p}_{\phi G}| 
=\Gamma_{J/\psi\to \omega G}/|{\bf p}_{\omega G}| $
due to the flavor-blind assumption. 
For $f_0^i$ decays into the same pseudoscalar meson pairs, such 
as $f_0^i\to K\bar{K}$, the ratio between these two partial 
decay widths becomes 
\be
\label{ozi-ratio}
R^{OZI}_i=\frac{\Gamma_{J/\psi\to\phi f_0^i\to\phi K\bar{K}}}
{\Gamma_{J/\psi\to\omega f_0^i\to\omega K\bar{K}}}
=\frac{| {\bf p}_{\phi i}|}{| {\bf p}_{\omega i} |}
\frac{[  x_i + y_i +\sqrt{2}r z_i ]^2}{2[ x_i  + r y_i + \sqrt{2} z_i ]^2} \ ,
\ee
which will be the focus of our discussions in this work. 
First, we point out that, 
apart from the momenta, $|{\bf p}_{\phi i}|$
and $|{\bf p}_{\omega i}|$, the energy scale does not explicitly appear in the 
above expression. The ratio $R^{OZI}_i$ will then be determined correlatively 
by the scalar mixing matrix elements and the doubly OZI violations ($r$). 
Also, we note that the mixing matrix elements contain the structure 
information about the scalars and are independent of the 
energy scales. In contrast, the doubly OZI violation parameter $r$ 
could depend on the energy scale. 
Therefore, 
it is natural to expect that the energy-dependence of $r$ would lead to 
the change of $R^{OZI}_i$ at different decay energies, 
or between, e.g., $J/\psi$ and $\Upsilon$ decays. 
Also, due to the correlation between $r$ and the mixing matrix elements, 
certain pattern for the energy-dependence of $R^{OZI}_i$ can be predicted.

As an example, to the limit of 
$r\to 0$, where the OZI rule applies, Eq.~(\ref{ozi-ratio})
 for the $f_0(1710)$ production will reduce to
\be
\label{no-r}
R^{OZI}_1\simeq \frac{| {\bf p}_{\phi 1}|}{2| {\bf p}_{\omega 1} |}
\frac{[  x_1 + y_1 ]^2}{[ x_1 + \sqrt{2} z_1 ]^2} \simeq 3 \ .
\ee
In the case of $J/\psi$ decays~\cite{close-zhao-f0}, 
this value is contradicted by the experimental data 
$R^{exp}_1\simeq 0.15$ from BES~\cite{bes-plb,bes-phi}, 
and 0.75 from PDG~\cite{pdg2004}. 
Substituting the experimental data into the above equation, 
one determines $r\simeq 2.2$ and 0.5, respectively. 
This suggests that there are important contributions from the OZI rule violations 
in the $J/\psi$ decays.

Although it may not be very surprising that there are significant 
OZI violation contributions in the scalar productions, 
it nontrivially implies that its presence and absence 
would lead to inversely changed values for $R_1^{OZI}$ as shown 
by Eqs.~(\ref{ozi-ratio}) and (\ref{no-r}). 
Note that though Eq.~(\ref{ozi-ratio}) is derived in the $J/\psi$ decays, 
it does not strongly depend on the masses of the $J/\psi$ meson except for 
$r$, of which the energy-dependence 
is not known~\footnote{Here, we neglect the trivial 
kinematic dependence arising from 
the mass differences between $\phi$ and $\omega$.}. 
In another word, by measuring the energy dependence of 
\be
\label{ratio-phi-omega}
R_i^{OZI}\equiv \frac{\Gamma_{3g(1S)\to \phi f_0^i}}
{\Gamma_{3g(1S)\to \omega f_0^i}}
\simeq 
\frac{[  x_i + y_i +\sqrt{2}r z_i ]^2}{2[ x_i  + r y_i + \sqrt{2} z_i ]^2}, 
\ee
one would have access to the energy evolution of 
the OZI violation contributions contained within $r$. 
As a result, the correlated information 
about the scalar structures can also be examined.

\begin{figure}
\begin{center}
\epsfig{file=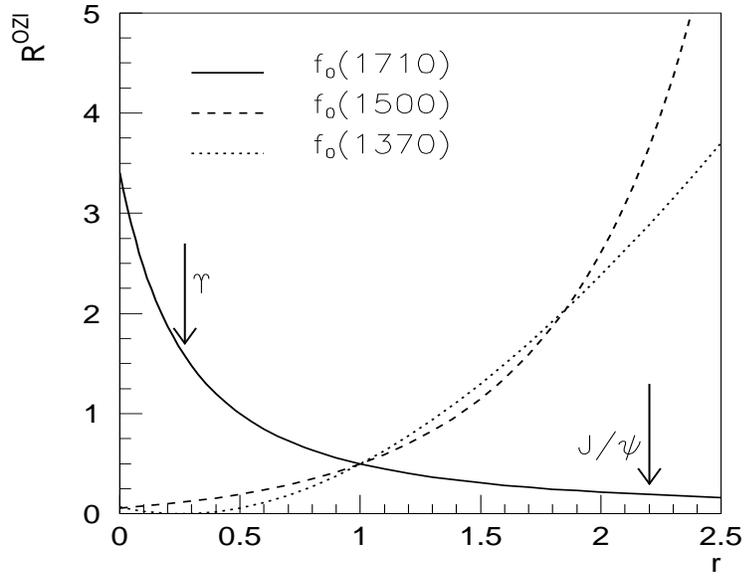, width=12cm,height=10.cm}
\caption{Energy evolution of $R^{OZI}_i$ in terms of parameter $r$. 
}
\protect\label{fig-2}
\end{center}
\end{figure}

Equation~(\ref{ratio-phi-omega}) allows us to 
investigate the evolution of $R^{OZI}_i$ in terms of a range of $r$, 
e.g. $r=0\sim 3$. In such a general context, $r\to 0$ will correspond to the 
regime where the OZI rule is well respected, 
while $r\sim 1$, to the region of strong OZI violations. 
We plot the ratio changes in Fig.~\ref{fig-2}.  
For the $J/\psi$ decays, $r\simeq 2.2$ 
suggests the significant breaking of the OZI rule 
as found in Ref.~\cite{close-zhao-f0}. 
This is the region that the small fraction of 
$R^{OZI}_1=0.15$ for the $f_0(1710)$ is observed. 
Furthermore, the ratios $R^{OZI}_2$ and $R^{OZI}_3$ 
for the $f_0(1500)$ and $f_0(1370)$ are much larger than 
unit. Hence, the branching ratio pattern observed by BES 
emerges.

In Fig.~\ref{fig-2}, in the limit of $r\to 0$, 
$R^{OZI}_i$ exhibits dramatical changes. In particular, 
it shows that the production branching ratio for 
$\phi f_0(1710)$ will become much larger than that for 
$\omega f_0(1710)$. Therefore, the ratio $R^{OZI}_1=0.15$ 
observed in the $J/\psi$ decays will be inversely changed 
to $R^{OZI}_1 > 1$.  
In fact, such an inverse change could happen even at moderate 
values for $r$. As shown by Fig.~\ref{fig-2}, 
the ratio $R^{OZI}_1=1$ corresponds to $r=0.5$. 
In brief, Fig.~\ref{fig-2} provides a guide for the change 
of the branching ratio fractions at different energies 
via the energy evolution of $r$. 
However, such an evolution trend 
will only become useful when the energy dependence of 
$r$ is quantified.  This requires an explicit 
relation between the doubly and singly OZI disconnected 
processes to be established. 
It will then enable us to
map $r$ in Fig.~\ref{fig-2} to a certain energy region, 
where the ratio $R^{OZI}_i$ can be predicted.

Theoretically, the energy evolution of $r$ can be investigated 
by calculating the relative strengths between the doubly disconnected 
process and the singly disconnected one as defined 
by Eq.~(\ref{para-r}).
However, due to the break-down of the perturbative character 
of the doubly disconnected processes, Fig.~\ref{fig-1}(c) 
cannot be regarded as Feynman diagram, but rather a topological 
one. In this sense, a pQCD calculation of $r$ becomes unreliable. 
We hence introduce a dynamical assumption based on sucessful 
phenomenological treatments available in the literature.  
We assume that the OZI doubly disconnected processes are dual 
with the intermediate meson exchange 
processes~\cite{lipkin,isgur-geiger,lipkin-zou}. 

\begin{figure}
\begin{center}
\epsfig{file=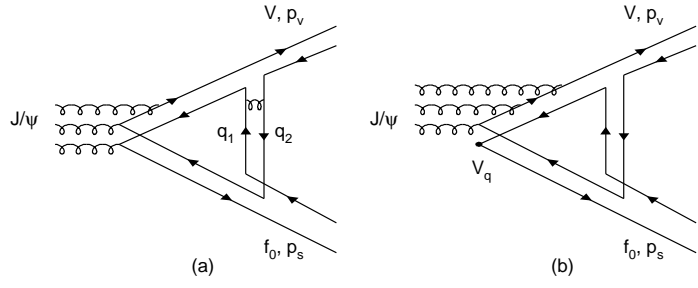, width=12cm,height=6.cm}
\caption{The OZI doubly disconnected processes in terms of 
intermediate virtual meson exchanges. 
}
\protect\label{fig-3}
\end{center}
\end{figure}

In the picture of Geiger and Isgur's intermediate virtual meson exchanges, 
the doubly disconnected diagram can be illustrated by 
two topological diagrams, Fig.~\ref{fig-3}(a) and (b). 
Qualitatively, such processes will be suppressed relative 
to the singly disconnected ones at higher energies. 
In Fig.~\ref{fig-3}(a), the two colored quark-antiquark pairs must 
exchange a colored gluon to keep overall color-singlet. 
In the case that the created quark pairs scatter out 
to form the final state $V$ and $f_0^i$, there must need  
large momentum transfers between quark line $q_1$ and $q_2$. 
Nevertheless, the momentum difference between $q_1$ and $q_2$ 
will increase with the increasing masses of the initial quarkonia 
since they must be recoiled back to the directions of their 
original quark (antiquark) partners to form the mesons in the final state. 
As a consequence, the exchanged gluon will become very 
hard and the transition amplitude will be suppressed 
at higher masses. One can also understand this as 
the suppression from the wavefunction distribution 
of the virtual meson formed by $q_1$ and $q_2$. 
Higher energies will correspond to the wavefunction tail 
at shorter distances, which is far away from the 
large wavefunction density determined by the typical 
size of meson. 

In Fig.~\ref{fig-3}(b), the color-singlet quark-antiquark pair of the scalar meson 
is created and recoiled to the backward direction. 
In the virtual meson exchange model, as shown by Fig.~\ref{fig-3}(b), 
the quark-antiquark pairs must correlate with each other to form 
the virtual mesons. Therefore, at least at one vertex, e.g. $V_q$, 
which is the $^3P_0$ production vertex, the quark line 
must be recoiled back approximately to the momentum direction
of the final antiquark. Because of such a large momentum transfer 
at vertex $V_q$, the $^3P_0$ mechanism is suppressed, which means 
that the transition via Fig.~\ref{fig-3}(b) will decrease 
with the increasing energies, i.e. in the decay of heavier quarkonia. 

In principle, 
by summing over all the possible meson exchange loops, 
the meson exchange mechanisms 
would provide the same results for the doubly OZI disconnected processes. 
However, difficulty will arise due to possible contributions from 
a large number of intermediate processes. 
A further reasonable assumption then is to consider the 
dominant processes based on the available 
experimental information. 
As shown by PDG~\cite{pdg2004}, $J/\psi$ has the largest partial decay widths to 
$\rho\pi$, and $\omega$ meson has also strong couplings to $\rho\pi$. 
Meanwhile, for these $f_0$ states, the decay of $f_0^i\to \pi\pi$ 
is one of the most important channels, e.g. $f_0(1370)\to \pi\pi$. 
This makes the $\rho\pi$ intermediate 
exchange a leading contribution to $J/\psi\to \omega f_0^3$. 
Similarly, the intermediate $K^*\bar{K}+c.c.$ exchange should be also 
a dominant contribution to $J/\psi\to \phi f_0^1$: the partial decay 
width of $J/\psi\to K^*\bar{K}+c.c.$ is about one order of magnitude 
larger than $J/\psi\to \phi f_0^1$; the $\phi$ meson couples 
strongly to $K^*K$; and $K\bar{K}$ is the dominant decay channel 
for the $f_0(1710)$. 
Furthermore, note that 
the other hadronic decay channels of $J/\psi$ 
are almost the same order of magnitude 
as $J/\psi\to \omega f_0(1710)$, or just one magnitude larger than 
$J/\psi\to \phi f_0(1710)$. Their rescatterings to final state $V f_0^i$ 
would be strongly suppressed due to 
their weak couplings to $V f_0^i$ and energy suppressions 
at the rescattering vertices.

\begin{figure}
\begin{center}
\epsfig{file=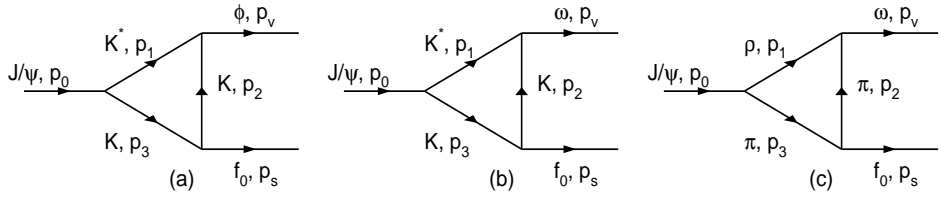, width=16cm,height=6.cm}
\caption{The OZI violations from intermediate $K^*K$ and $\rho\pi$ 
meson exchanges as dominant contributions to the doubly disconnected processes. 
}
\protect\label{fig-4}
\end{center}
\end{figure}

Following the meson exchange mechanisms, 
we calculate the meson loops in Fig.~\ref{fig-4}  
as a leading contribution to the doubly disconnected process.
Supposing the OZI violation effects are from the intermediate meson 
exchange processes, we can estimate the meson exchange contribution 
and compare it with the ``tree" diagram for $J/\psi\to V\bar{P}+c.c.$, 
e.g. $J/\psi\to K^* \bar{K}+c.c.$ 
and $J/\psi \to \rho\pi+c.c.$  
Since the $J/\psi V\bar{P}$ vertices are the same between the tree and loop 
transitions, the ratio between the meson-exchange loop 
and the tree process of $J/\psi\to V\bar{P}+c.c.$ will then highlight the 
OZI violations in this dynamical process. 
For the purpose of gaining information about the 
rough behavior of the energy-evolution of the doubly OZI processes, 
the energy-evolution of this ratio is sufficient to account for the 
energy-dependence of the doubly OZI violation transitions.

The transition ampliude can be written as 
\be
M_{fi}=-i\int\frac{d^4 p_2}{(2\pi)^4}T_v^\beta
\left(g^\lambda_\beta-\frac{p_{1\beta}p_1^\lambda}{p_1^2}\right)
T_{0\lambda} T_s\frac{F(p_2)}{a_1 a_2 a_3} \delta^4(p_0-p_v-P_s) \ ,
\ee
where the vertex functions are 
\bea
\label{vertex-func}
T_v^\beta &= & i\frac{g_v}{M_v}\epsilon^{\mu\nu\alpha\beta}
p_{v\mu}\epsilon_{f\nu}p_{2\alpha}, \nonumber\\
T_\lambda &=& i\frac{g_0}{M_0}
\epsilon_{\lambda\sigma\tau\delta}p_0^\sigma\epsilon_i^\tau p_3^\sigma , \nonumber\\
T_0 & =& ig_s M_s \ , 
\eea
and $a_1=p_1^2-m_1^2+i\epsilon$, $a_2=p_2^2-m_2^2 +i\epsilon$, and $a_3 = p_3^2-m_3^2 +i\epsilon$ 
are the denominators of the meson propagators. 
The coupling constants $g_v$, $g_s$ and $g_0$ can be determined independently 
in meson decays. For the $J/\psi$ hadronic decays, large branching ratios 
for $J/\psi\to K^*\bar{K} + c.c.$ and $ \rho\pi +c.c.$ imply that 
the meson loop transitions may have significant contributions to 
those decays where the final state mesons have also large couplings 
to the exchanged mesons. This is indeed the case for 
$J/\psi\to V f_0(1710)$, where the couplings for $\phi K^*K$, $\omega K^*K$, 
and $f_0(1710)K\bar{K}$ are significantly large. 
Therefore, this simple argument will allow us to consider only the 
dominant meson exchange loops in the calculations.

In the above equation, the coupling $g_0$ can be determined by the 
decays of 
$J/\psi \to K^*\bar{K}+c.c.$ [Fig.~\ref{fig-4}(a) and (b)], 
or $J/\psi\to \rho\pi+c.c.$ [Fig.~\ref{fig-4}(c)], of which 
large branching ratios are observed in experiment, e.g. 
\be
g_0^2=\frac{12\pi M_0^2}{|{\bf p}_1|^3} \Gamma^{exp}_{J/\psi\to V\bar{P}} \ ,
\ee
where $\Gamma^{exp}_{J/\psi\to K^*\bar{K}+c.c.}= (9.2\pm 0.8)\times 10^{-3}$ 
and 
$\Gamma^{exp}_{J/\psi\to \rho\pi+c.c.}=(1.27\pm 0.09)\% $
are from the estimate of Particle Data Group~\cite{pdg2004}.

For  $f_0\to K\bar{K}$ in the $f_0$ c.m. system, 
the coupling constant can be derived via 
\be
g_s^2=\frac{8\pi}{|{\bf k}|}\Gamma^{exp}_{f_0\to K\bar{K}} \ ,
\ee
which is consistent with the studies of $f_0\to PP$ 
in the determination of the 
mixing matrix elements~\cite{close-zhao-f0}; 
$|{\bf k}|$ is the magnitude of the three momenta carried by the final-state 
kaon (anti-kaon).
With the estimate of $b.r._{f_0\to K\bar{K}}=0.60$~\cite{close-zhao-f0}, 
and $\Gamma_T(1710)=140$ MeV~\cite{pdg2004}, we have 
$\Gamma^{exp}_{f_0\to K\bar{K}}=84$ MeV; 
the coupling $g_s$ can then be determined.

We determine the $VVP$ couplings, i.e., 
$g_{\phi K^* K}$, $g_{\omega K^*K}$, and $g_{\omega \rho\pi}$, 
in the SU(3)-flavor-symmetry limit. Since $\omega$ and $\phi$
are almost ideally mixed, the SU(3) symmetry leads to 
\be
g_{\omega \rho\pi}=3g_{\omega \rho^0\pi^0}, \ \ 
g_{\phi K^* K}=2\sqrt{2} g_{\omega \rho^0\pi^0}, \ \ 
g_{\omega K^*K}=2g_{\omega \rho^0\pi^0},
\ee
where $g^2_{\omega \rho^0\pi^0}\simeq 84$ is determined 
in vector meson dominance (VMD) model 
in $\omega\to \pi^0 e^+ e^-$~\cite{pdg2004}.

We apply the Cutkosky rule to the loop calculation, where the mesons 
from the decay vertex are assumed to be on-shell, i.e. 
$p_1^2=m_1^2$ and $p_3^2=m_3^2$. 
The transition amplitude then reduces to 
\be
\label{loop}
M_{fi}=-\frac{ig_0 g_v g_s}{64\pi^2}[p_v\cdot p_0\epsilon_f\cdot \epsilon_i
-p_v\cdot \epsilon_i p_0\cdot \epsilon_f] 
\frac{M_s |{\bf p_3}|}{M_0^2 M_v} {\cal S} \ ,
\ee
where
\be
\label{int-0}
{\cal S}=\int d\Omega 
\frac{(p_s-p_3)^2 F((p_s-p_3)^2)}{(p_s-p_3)^2-m_2^2} \ ,
\ee
is the reduced loop integral after applying 
the on-shell condition for the integrand; 
The momentum conservation, $p_2= p_s-p_3$, hence leads to the 
angular dependence of the integrand in the above equation since
$p_2^2=(p_s-p_3)^2=M_s^2+m_3^2-2E_s E_3 +2|{\bf p}_s||{\bf p}_3|\cos\theta $. 
The form factor
 $F((p_s-p_3)^2)$ is included to take into account the off-shell
effects for the exchanged meson in the final state interactions.

In the loop transition, the polarizations of the initial-final 
state vector mesons, longitudinal-transverse and 
vice versa (TL), will not contribute to the integrals. 
Also, polarization of longitudinal-longitudinal (LL) vanishes in 
the tree transitions. We hence take the transverse-transverse (TT) 
ratio between the loop and tree transitions to define 
\be 
r_c\equiv -\frac{g_v g_s}{64\pi^2}
\frac{M_s E_v}{M_0 M_v} {\cal S} \ ,
\ee
where $E_v$ is the energy of the final state vector meson, and $r_c$ 
reflects the energy evolution of the meson loop transitions. 
For the loop transition, the LL component compared to the LL ones 
is $M_v/E_v$ suppressed at higher energies, i.e., $M_0\to \infty$. 
Therefore, $r_c$ is a good quantity for describing the energy evolutions of 
the loop transitions relative to the tree ones. 
A feature arising from $r_c$ 
is that $E_v/M_0\to 1/2$ in the limit of 
$M_0\to \infty$. Therefore, for sufficiently large $M_0$, 
the energy evolution of $r_c$ will be determined by ${\cal S}$.

The meson loop integration of Eq.~\ref{int-0} can be rewritten as 
\bea
{\cal S}&=&-2\pi\frac{B}{A}\int d\cos\theta \frac{1+A\cos\theta}{1+B\cos\theta}
F((p_s-p_3)^2) \ ,
\eea
where 
\bea
A &\equiv & 2|{\bf p}_s||{\bf p}_3|/(M_s^2+m_3^2-2E_s E_3) \nonumber\\
B &\equiv & 2|{\bf p}_s||{\bf p}_3|/(M_s^2+m_3^2-2E_s E_3-m_2^2) \ .
\eea
For the following situations with explicit considerations of 
the form factors, the integration can be worked out
analytically:

i) Without form factor: $F((p_s-p_3)^2)=1$

The integration gives
\be
\label{int-1}
{\cal S}_a=2\pi\left[2-\frac{B-A}{AB}\ln\frac{1-B}{1+B}\right] \ .
\ee

ii) Monopole form factor: 
$F((p_s-p_3)^2)=(\Lambda_1^2-m_2^2)/(\Lambda_1^2-p_2^2)$

Define
\bea 
C &\equiv & 2|{\bf p}_s||{\bf p}_3|/(\Lambda_1^2-(M_s^2+m_3^2-2E_s E_3))\nonumber\\
\lambda &\equiv & (\Lambda_1^2-m_2^2)/(\Lambda_1^2-(M_s^2+m_3^2-2E_s E_3)) \ ,
\eea
we have 
\be
\label{int-2}
{\cal S}_b=2\pi\lambda \left[\frac{1}{C}\ln\frac{1+C}{1-C}
+\frac{A-B}{A(B-C)}\ln\frac{(1-B)(1+C)}{(1+B)(1-C)}\right] \ .
\ee

iii) Dipole form factor: 
$F((p_s-p_3)^2)=[(\Lambda_2^2-m_2^2)/(\Lambda_2^2-p_2^2)]^2$

The integration gives
\be
\label{int-3}
{\cal S}_c=2\pi\lambda^2 \left\{\frac{2}{1-C^2}
+\frac{A-B}{A}\left[\frac{2C}{(B-C)(1-C^2)}
+\frac{B}{(B-C)^2}\ln\frac{(1-B)(1+C)}{(1+B)(1-C)}\right]\right\} \ .
\ee

In (ii) and (iii) parameters $\Lambda_1$ and $\Lambda_2$ have a range of 
$1\sim 2$ GeV, which are commonly adopted 
in the literature~\cite{li-bugg-zou}. 
In general, the adopted value for $\Lambda_1$ is smaller than 
that for $\Lambda_2$. 
In this calculation, we adopt $\Lambda_1=1.2$ GeV, and $\Lambda_2=1.9$ GeV, 
since they lead to the same results for ${\cal S}_b$ and ${\cal S}_c$ 
in $J/\psi$ decays. If the same value for $\Lambda_1$ and $\Lambda_2$ 
is adopted, we find ${\cal S}_c/{\cal S}_b\simeq 0.33 \sim 0.55$ within 
the range of $1.2 \sim 1.9$ GeV for the cut-off energy.

In Table~\ref{tab-1}, the ratio $r_c$ is calculated for 
$J/\psi\to K^*\bar{K}+c.c.\to \phi f_0^i$. 
Without form factor, the loop transition accounts for 
large fraction of the $J/\psi\to K^*\bar{K}+c.c.$ decay, 
and obviously overestimates the branching ratios for $J/\psi\to V f_0^i$. 
In contrast, the inclusion of form factors will significantly 
suppress the loop transition strengths. 
Roughly to say, the loop transition has the largest 
contributions to the production of $f_0(1710)$. 
This can be explained by the large $f_0(1710)K\bar{K}$ couplings. 
Interestingly, 
the $K^*K$ loop contributions to the branching ratio 
of $J/\psi\to\phi f_0(1710)\to \phi K\bar{K}$ 
have the same order of magnitude as the experimental data. 
This directly shows that large OZI violations occur via the 
loop transitions. Also, the suppressed loop contributions 
in the production of $f_0(1500)$ and $f_0(1370)$ are consistent 
with the experimental results~\cite{bes-phi} and calculations presented 
in Ref.~\cite{close-zhao-f0}. 

For $J/\psi\to \omega f_0^i$, in addition to the $K^*K$ loop transition, 
the $\rho\pi$ loop may also be significantly large. 
In Table~\ref{tab-2},  the loop contributions from 
$J/\psi\to K^*\bar{K}+c.c.\to\omega f_0^i$ are presented. 
Again, we obtain a relatively larger branching ratio 
for the $f_0(1710)$ than for the $f_0(1500)$ and $f_0(1370)$. 
Compared with the experimental data~\cite{bes-plb}, where 
$\Gamma_{J/\psi\to\omega f_0(1710)} > \Gamma_{J/\psi\to\phi f_0(1710)}$, 
it shows that the loop transition contributes slightly larger 
in $J/\psi\to\phi f_0(1710)$ than in $J/\psi\to\omega f_0(1710)$. 
This originates from $g_{\phi K^* K}>g_{\omega K^*K}$ in the loop.

The story changes in the $\rho\pi$ loop transition. 
As presented in Table~\ref{tab-3}, this process has sizeable 
contributions to the $f_0(1500)$ and $f_0(1370)$ productions, 
but is negligibly small in the production of $f_0(1710)$. 
This suggests much more important roles are played by the $\rho\pi$ 
loop transitions for those two states. 

Although we do not present the estimate of the contributions from 
the $\rho\pi$ loop transition to $J/\psi\to \rho\pi+c.c.\to \phi f_0^i$, 
it is worth noting that the $\phi\rho\pi$ coupling should not be small. 
As quoted by PDG~\cite{pdg2004}, the branching ratio for 
$\phi\to \rho\pi+ \pi^+\pi^-\pi^0$ is about 15.4\%. 
Supposing a large fraction of the decays is via $\phi\to \rho\pi$, 
it will give a large coupling, and  
naturally lead to large branching ratios for the $f_0(1370)$ 
and $f_0(1500)$ in the $\phi\pi\pi$ channel.

It should be noted that the loop transitions 
could contribute more to the branching ratios. 
Due to the on-shell approximation applied, we have only included 
the imaginary part of the  transition amplitudes in the calculation, 
and neglected contributions from the real part. 
More rigorous approaches would also consider the latter via e.g. disperse 
relation~\cite{cheng}. For the purpose of illustrating 
large OZI violations via the loop transitions, and investigating their 
energy evolution, it is sufficient to only consider the 
imaginary amplitudes from the on-shell approximation.

To relate the above loop transitions to the OZI violation parameter $r$,
we first note that the ratio $r$ does not equivalent to 
$r_c$. In the definition of $r_c$, the tree coupling is given by 
$J/\psi\to K^*\bar{K}+c.c.$, while for $r$, the tree coupling is from the 
singly disconnected processes. 
Due to lack of constraints on the singly disconnected processes 
of $J/\psi\to \phi f_0$, e.g. 
Fig.~\ref{fig-1}(a) and (b), we cannot directly derive $r$. 
However, notice that the major difference between $r$ and $r_c$ 
is the different coupling at the $J/\psi$ decay vertex. We can then estimate:  
\be
\label{energy-dep}
\frac{r(M_\Upsilon)}{r(M_{J/\psi})} 
\simeq \frac{r_c(M_\Upsilon)}{r_c(M_{J/\psi})} \ .
\ee
Such a relation must be a robust one since we do not include all the 
intermediate meson exchange contributions. However, as we addressed in the 
previous part, the dominant meson-loop contributions are from the 
processes that the couplings at those three vertices are strong. 
In this sense, the neglect of other meson exchange contributions 
should not result in magnitude uncertainties in the above estimate. 
As confirmed by the calculations that 
the loop transitions are the dominant contributions 
to the OZI violation processes and have the same order of magnitude as 
the singly disconnected processes,
 we adopt $r(M_{J/\psi})=2.2$ 
from Ref.~\cite{close-zhao-f0} to derive 
$r(M_\Upsilon)\simeq 0.26\sim 0.28$, 
which indicates that the loop transition amplitude is indeed suppressed 
at the $\Upsilon$ mass region.
The variation is from calculations for 
the $K^*K$ and $\rho\pi$ loops, respectively.

Mapping $r(M_\Upsilon)\simeq 0.27\pm 0.01$ to Fig.~\ref{fig-2} (see the arrows 
in the figure), 
we find that $R^{OZI}_1\simeq 1.67\pm 0.04$ in contrast with 
$R^{OZI}_1=0.15$ at $r(M_{J/\psi})=2.2$. As shown by 
the curves in Fig.~\ref{fig-2} between $r=0.27 -2.2$, 
$R^{OZI}_i$ experiences a dramatical change, which 
suggests a measurable effect 
from the possible OZI violation processes in the energy evolutions. 
In the small $r$ region, 
where the OZI rule applies, the production widths for 
$f_0(1500)$ and $f_0(1370)$ in the $\phi$ channel will 
be significantly suppressed in comparison with the $\omega$ channel. 
At the $\Upsilon$ energy, i.e. $r\simeq 0.27$, 
the ratios $R^{OZI}_2$ and $R^{OZI}_3$ are also  
inversely changed in contrast with the values derived at  
the $J/\psi$ mass region. 

\begin{figure}
\begin{center}
\epsfig{file=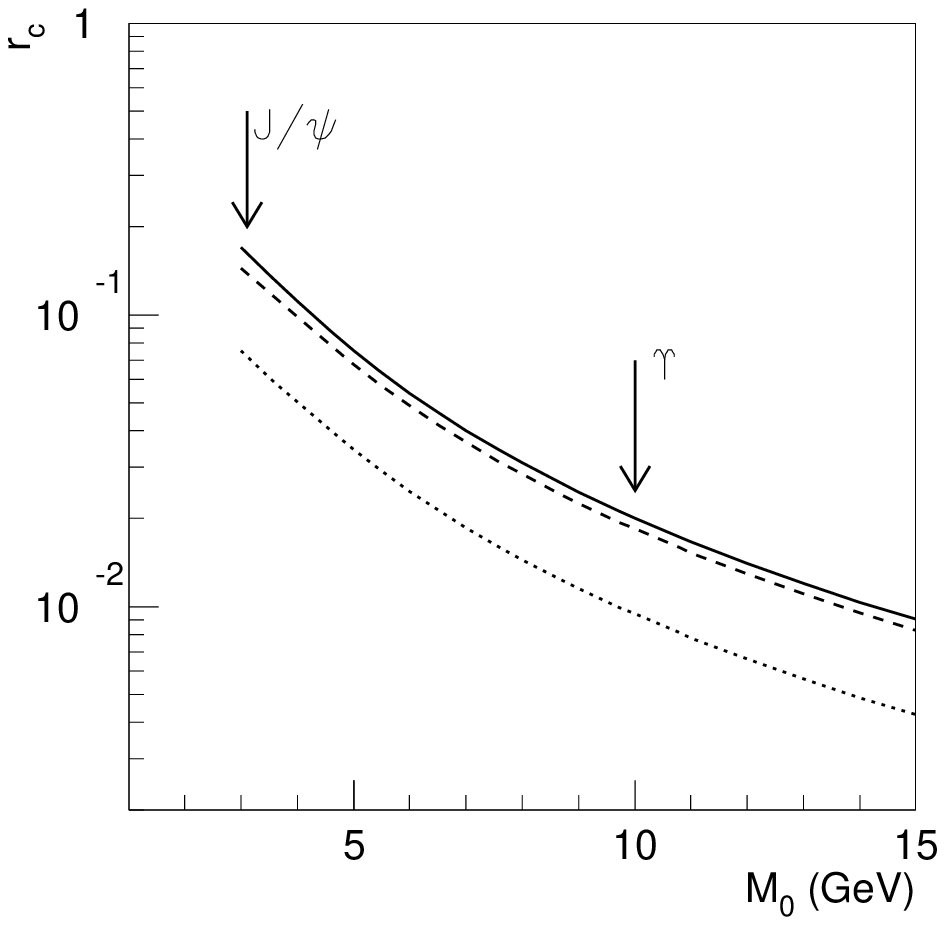, width=10cm,height=10cm}
\caption{The energy evolution of the ratio between the 
dominant meson loop transitions and the corresponding tree transitions. 
The solid curve is for $K^*K$ loop in the $\phi f_0(1710)$ channel, while the
dashed and dotted curves are for the $K^*K$ and $\rho\pi$ loops 
in the $\omega f_0(1710)$ channels. 
}
\protect\label{fig-5}
\end{center}
\end{figure}

The energy evolution of $r_c$ in those three loops 
are presented in Fig.~\ref{fig-5}. 
Exponential drop can be seen at small $M_0$ region, 
and it slightly levels off above 7 GeV. This indicates explicitly 
that the contributions from the loop transitions 
will decrease much faster than the tree transitions. 
This is consistent with the qualitative analyses 
made earlier. 
One can also see that the three $r_c$'s have similar 
behaviors with the increasing energies. This eventually justifies 
the above simple relation as a general 
results for relating $r$ to $r_c$ through 
the dominant loop transitions.

Undoubtedly, the measurement of $R^{OZI}_i$ 
in the $\Upsilon$ decays also examines the glueball-$Q\bar{Q}$ 
mixing scheme proposed for the scalar mesons. 
As presented in Fig.~\ref{fig-2}, the dramatic 
change of the relative positions of  
$R^{OZI}_i$ for these three $f_0$ states also depend on the 
mixing matrix elements. For example, 
in addition to the measurement of $R^{OZI}_1$ for the 
$f_0(1710)$ in $\Upsilon\to V f_0(1710)\to V K\bar{K}$, 
it is also interesting to measure 
$R^{OZI}_3$ for the $f_0(1370)$ in 
$\Upsilon\to Vf_0(1370)\to V\pi\pi$. The values for 
$R^{OZI}_1$ and $R^{OZI}_3$ if match the pattern 
of Fig.~\ref{fig-2}, would be a strong support for 
the glueball-$Q\bar{Q}$ mixings. 
The $R^{OZI}_2$ for the $f_0(1500)$ can also be investigated though
difficulty may arise from the small branching ratios
for both $\Upsilon\to \phi f_0(1500)$ and $\Upsilon\to \omega f_0(1500)$. 
Although, high statistics in experiment should be required
to isolate these two decay channels, the energy-evolution trends
of $R^{OZI}_i$ seem to provide a way to disentangle 
the correlation between the OZI-rule violation and 
the structure of those scalar mesons.

In summary, we discussed a possible way to determine 
the scalar structures by clarifying the role 
played by the OZI-rule violation. The latter was correlated 
with the glueball-$Q\bar{Q}$ mixings in $J/\psi\to V f_0^i$. 
Since the flavor wavefunctions for $\omega$ and $\phi$ are almost ideally mixed, 
the decay channels into $\omega$ and $\phi$ in association with 
the scalar mesons, respectively, serve as a flavor filter 
for probing the $Q\bar{Q}$ contents of the scalars. 
This allows us to separate out the doubly OZI 
disconnected processes,  of which the effects can be 
measured by the branching ratio fractions between 
$\phi f_0^i$ and $\omega f_0^i$, i.e. $R^{OZI}_i$. 
Since the energy evolution of $R^{OZI}_i$ is mostly determined 
by the energy evolution of the doubly disconnected processes 
relative to the singly disconnected ones, 
the suppression of the doubly disconnected process 
at higher energies, e.g. in $\Upsilon$ decays,  
will lead to dramatic changes to  
$R^{OZI}_i$ with certain patterns. Observation of such a change 
will provide direct information 
about the scalar meson structures.

Although at this moment, experimental data for 
$\Upsilon\to V f_0^i$ are unavailable, clear signals for $f_0(1710)$ 
have been reported in $\Upsilon\to \gamma f_0(1710)\to \gamma K\bar{K}$~\cite{pdg2004}. 
With an increasing statistics, one may have access to 
the above mentioned channels at future facilities for heavy 
quarkonium physics~\cite{dytman,qwg}.

The authors thank F.E. Close for many useful discussions and comments on this 
work. This work is supported,
in part, by grants from
the U.K. Engineering and Physical
Sciences Research Council Advanced Fellowship (Grant No. GR/S99433/01),
and the Institute of High Energy, Chinese Academy of Sciences.


\begin{table}[ht]
\begin{tabular}{c|c|c|c}
\hline
 & \ \ \  \ \ \ $f_0(1710)$ \ \ \ \ \ \ 
 & \ \ \ \ \ \  $f_0(1500)$ \ \ \  \ \ \ 
 & \ \ \ \ \ \  $f_0(1370)$ \ \ \ \ \ \   \\[1ex]
 & (i) \ \ \ \ \ (ii) \ \ \ \ \ (iii) 
 & (i) \ \ \ \ \ (ii) \ \ \ \ \ (iii)
 & (i) \ \ \ \ \ (ii) \ \ \ \ \ (iii)  \\[1ex]
\hline
$r_c(M_{J/\psi})$ 
& 0.550 \ \ \ 0.163 \ \ \ 0.165
& 0.196 \ \ \ 0.056 \ \ \ 0.057
& 0.149 \ \ \ 0.041 \ \ \ 0.042 \\[1ex]
$b.r.(J/\psi\to\phi f_0) $
& 19.31 \ \ \ 1.71 \ \ \ 1.73
& 2.95 \ \ \ 0.24 \ \ \ 0.24
& 1.83 \ \ \ 0.14 \ \ \ 0.15 \\[1ex]
$b.r.(J/\psi\to\phi f_0\to \phi K\bar{K}) $ 
& {\bf 11.59 \ \ \ 1.02 \ \ \ 1.04 }
& 2.53 \ \ \ 0.02 \ \ \ 0.02 
& 0.04 \ \ \ 0.00 \ \ \ 0.00 \\[1ex]
$b.r.(exp)$ & $(2.0\pm 0.7)$ & $(0.8\pm 0.5)$ & $(0.3\pm 0.3)$ \\[1ex]
\hline
\end{tabular}
\caption{ The intermediate $K^*K$ meson exchange contributions to the decay of 
$J/\psi\to\phi f_0\to \phi K\bar{K}$ for 
three cases: (i) {\it without} form factors; 
(ii) {\it with} monopole form factors; 
and (iii) {\it with} dipole form factors. 
The channel to which the loop transition has significant contributions 
is highlighted by bold characters. 
The numbers in the round brackets 
are data for $J/\psi\to\phi f_0\to \phi K\bar{K}$ 
from BES~\cite{bes-phi}. 
The branching ratios are in a unit of 
$10^{-4}$.}
\label{tab-1}
\end{table}

\begin{table}[ht]
\begin{tabular}{c|c|c|c}
\hline
 & \ \ \  \ \ \ $f_0(1710)$ \ \ \ \ \ \ 
 & \ \ \ \ \ \  $f_0(1500)$ \ \ \  \ \ \ 
 & \ \ \ \ \ \  $f_0(1370)$ \ \ \ \ \ \   \\[1ex]
 & (i) \ \ \ \ \ (ii) \ \ \ \ \ (iii) 
 & (i) \ \ \ \ \ (ii) \ \ \ \ \ (iii)
 & (i) \ \ \ \ \ (ii) \ \ \ \ \ (iii)  \\[1ex]
\hline
$r_c(M_{J/\psi})$ 
& 0.479 \ \ \ 0.139 \ \ \ 0.141
& 0.172 \ \ \ 0.047 \ \ \ 0.049
& 0.131 \ \ \ 0.034 \ \ \ 0.036 \\[1ex]
$b.r.(J/\psi\to\omega f_0) $
& 16.47 \ \ \ 1.39 \ \ \ 1.43
& 2.39 \ \ \ 0.18 \ \ \ 0.19
& 1.46 \ \ \ 0.10 \ \ \ 0.11 \\[1ex]
$b.r.(J/\psi\to\omega f_0\to \omega K\bar{K}) $ 
& {\bf 9.88 \ \ \ 0.84 \ \ \ 0.86 }
& 0.21 \ \ \ 0.02 \ \ \ 0.02 
& 0.03 \ \ \ 0.00 \ \ \ 0.00 \\[1ex]
$b.r.(exp)$ & $(13.2\pm 2.6)$ & $\cdots$ & $\cdots$ \\[1ex]
\hline
\end{tabular}
\caption{ The intermediate $K^*K$ meson exchange contributions to the decay of 
$J/\psi\to\omega f_0\to \omega K\bar{K}$ presented in a way 
similar to Table~\ref{tab-1}. The numbers in the round brackets 
are data for $J/\psi\to\omega f_0\to \omega K\bar{K}$ 
from BES~\cite{bes-plb}, and the dottes denote the unavailablity of the 
experimental data. The branching ratios are in a unit of 
$10^{-4}$.}
\label{tab-2}
\end{table}
 
\begin{table}[ht]
\begin{tabular}{c|c|c|c}
\hline
 & \ \ \  \ \ \ $f_0(1710)$ \ \ \ \ \ \ 
 & \ \ \ \ \ \  $f_0(1500)$ \ \ \  \ \ \ 
 & \ \ \ \ \ \  $f_0(1370)$ \ \ \ \ \ \   \\[1ex]
 & (i) \ \ \ \ \ (ii) \ \ \ \ \ (iii) 
 & (i) \ \ \ \ \ (ii) \ \ \ \ \ (iii)
 & (i) \ \ \ \ \ (ii) \ \ \ \ \ (iii)  \\[1ex]
\hline
$r_c(M_{J/\psi})$ 
& 0.236 \ \ \ 0.072 \ \ \ 0.078
& 0.494 \ \ \ 0.143 \ \ \ 0.156
& 0.565 \ \ \ 0.159 \ \ \ 0.175 \\[1ex]
$b.r.(J/\psi\to\omega f_0) $
& 5.26 \ \ \ 0.49 \ \ \ 0.57
& 25.87 \ \ \ 2.18 \ \ \ 2.59
& 35.84 \ \ \ 2.85 \ \ \ 3.43 \\[1ex]
$b.r.(J/\psi\to\omega f_0\to \omega \pi\pi) $ 
& 0.35 \ \ \ 0.03 \ \ \ 0.04 
& {\bf 9.03 \ \ \ 0.76 \ \ \ 0.90 }
& {\bf 7.17 \ \ \ 0.57 \ \ \ 0.69 } \\[1ex]
$b.r.(exp)$ & $\cdots$ & $\cdots$ & $\cdots$ \\[1ex]
\hline
\end{tabular}
\caption{ The intermediate $\rho\pi$ meson exchange contributions 
to the decay of 
$J/\psi\to\omega f_0\to \omega \pi\pi$ presented in a way similar to 
Table~\ref{tab-1}. 
The dottes denote the unavailablity of the 
experimental data. The branching ratios are in a unit of 
$10^{-4}$.}
\label{tab-3}
\end{table}


\begin{thebibliography}{99}

%
\bibitem{bes-phi}
M.~Ablikim {\it et al.}  [BES Collaboration],
arXiv:hep-ex/0411001.
%
%
\bibitem{bes-plb}
M.~Ablikim {\it et al.}  [BES Collaboration],
Phys.\ Lett.\ B {\bf 603}, 138 (2004)
[arXiv:hep-ex/0409007].
%
\bibitem{bes-f0(1500)} M.~Ablikim {\it et al.}  [BES Collaboration],
Phys.\ Lett.\ B {\bf 598}, 149 (2004)
[arXiv:hep-ex/0406038].
%
\bibitem{jin} S. Jin, Plenary talk at 
32nd International Conference on High-Energy Physics (ICHEP2004), 
Beijing (2004). 
%
\bibitem{scalar-meson-1} F.E. Close and N.A. Tornqvist,
J. Phys. {\bf G 28}, R249 (2002).
%
\bibitem{close-amsler} F.E. Close and C. Amsler, 
Phys.\ Lett.\ B {\bf 353}, 385 (1995); Phys. Rev. {\bf D53}, 295 (1996).
%
\bibitem{close-ichep04} F.E. Close, 
 Plenary talk at 32nd International Conference on High-Energy Physics 
 (ICHEP2004), 
Beijing (2004) [archive:hep-ph/0411396].
%
\bibitem{mp} C. Morningstar and M. Peardon,
	Phys. Rev. D {\bf 56}, 4043 (1997).
%
\bibitem{ukqcd} G. Bali {\it et al.}, UKQCD Collaboration,
	Phys. Lett. {\bf B 309}, 378 (1993).
%
\bibitem{chanowitz} M.S. Chanowitz, 
Proc. 6th Int. Workshop on $\gamma\gamma$ Collisions, Edited by R. Lander 
(World Scentific, Singapore, 1984).
%
\bibitem{narison} S. Narison,
	Nucl. Phys. {\bf B 509}, 312 (1998).
%
\bibitem{cfl} F.E. Close, G.R. Farrar, and Z. Li,
        Phys. Rev. D {\bf 55}, 5749 (1997).
%
\bibitem{bugg} D.V. Bugg, M. Peardon, and B.S. Zou,
Phys. Lett. {\bf B 486}, 49 (2000). 
%
\bibitem{close-kirk} F.E. Close and A. Kirk, 
Phys.\ Lett.\ B {\bf 483}, 345 (2000).
%
\bibitem{close-zhao-f0} F.E. Close and Q. Zhao,
Phys. Rev. D {\bf 71}. 094022 (2005) [arXiv:hep-ph/0504043]; 
%
Phys.\ Lett.\ B {\bf 586}, 332 (2004)
[arXiv:hep-ph/0402090].
%
\bibitem{ozi} S. Okubo, Phys. Lett. {\bf 5}, 1975 (1963); 
G. Zweig, in {\it Development in the Quark Theory of Hadrons}, 
edited by D.B. Lichtenberg and S.P. Rosen (Hadronic Press, Massachusetts, 1980);
J. Iizuka, Prog. Theor. Phys. Suppl. {\bf 37}, 38 (1966).
%
\bibitem{seiden} A. Seiden, H. F.-W. Sadrozinski, and H. E. Haber, 
	Phys. Rev. D {\bf 38}, 824 (1988).
%
\bibitem{lipkin} H.J. Lipkin, 
Phys. Rev. Lett. {\bf 13}, 590 (1964); {\bf 14}, 513 (1965); 
Phys. Rep. {\bf 8C}, 173 (1973); 
Nucl. Phys. {\bf B244}, 147(1984); {\bf B291}, 720 (1987); 
Phys. Lett. {\bf B179}, 278 (1986). 
%
\bibitem{isgur-geiger} P. Geiger and N. Isgur, 
	Phys. Rev. D {\bf 47}, 5050 (1993). 
%
\bibitem{lipkin-zou} H.J. Lipkin and B.S. Zou, 
	Phys. Rev. D {\bf 53}, 6693 (1996).
%
\bibitem{locher-lu-zou}
  M.~P.~Locher, Y.~Lu and B.~S.~Zou,
  Z.\ Phys.\ A {\bf 347}, 281 (1994)
  [arXiv:nucl-th/9311021].
%
\bibitem{pdg2004} S. Eidelman {\it et al.}  (Particle Data Group),
Phys. Lett. {\bf B 592}, 1 (2004).
%
\bibitem{li-bugg-zou} X.Q. Li, D.V. Bugg, and B.S. Zou, 
	Phy. Rev. D {55}, 1421 (1997). 
%
\bibitem{cheng} H.-Y. Cheng, C.-K. Chua, and A. Soni,
	Phys. Rev. D {\bf 71}, 014030 (2005).
%
\bibitem{dytman} S.A. Dytman {\it et al.} [CLEO Collaboration],
	hep-ex/0307035.
%
\bibitem{qwg} N. Brambilla {\it et al.}, hep-ph/0412158.
%
%
%
%
%
%

\end{thebibliography}
\end{document}